# The Structure and Evolution of Lyman-alpha Forest Clouds in the Minihalo Model

Avery Meiksin[1]
Canadian Institute for Theoretical Astrophysics

## ABSTRACT

Results are presented for one-dimensional numerical hydrodynamics computations of the structure and evolution of Ly$\alpha$ forest clouds gravitationally confined by dark matter minihalos. The clouds are developed from linear perturbations at high redshift and exposed to either a QSO- or galaxy-dominated metagalactic radiation field at moderate redshifts. The baryonic component of the clouds is assumed to be composed of hydrogen and helium in cosmic abundance. While the emphasis is on spherical systems, slab symmetry is also considered. The effects of photoionization heating and losses from radiative atomic processes and from Compton cooling are included. Three zones may be identified in a collapsed cloud: (1) a quasi-hydrostatic core in thermal equilibrium, (2) a nonhydrostatic intermediate zone out of thermal equilibrium, and (3) a cosmological accretion layer joining onto the Hubble expansion. Most of the measured Ly$\alpha$ forest column densities arise in the intermediate zone. The development of the core would result in a flattening in the column density distribution near $\log N_{\rm HI} \sim 15-16$. The cloud diameters corresponding to a column density of $10^{14}\,{\rm cm}^{-2}$ lie in the range $10-60\,{\rm kpc}$, while systems with $N_{\rm HI} > 10^{15}\,{\rm cm}^{-2}$ have diameters smaller than 10 kpc. Because of a postphotoionization wind, the ratio of central baryon to dark matter density in the clouds generally lies below the cosmic value, although contraction toward hydrostatic equilibrium increases the ratio with time. Systems with circular velocities exceeding $v_c \sim 50\,{\rm km\,s^{-1}}$ result in clouds which contract until they become Jeans unstable and collapse. The critical column density for collapse is $10^{17}-10^{18}\,{\rm cm}^{-2}$.

A minimum Doppler parameter of $b \sim 25\,{\rm km\,s^{-1}}$ is found for absorption lines with column densities in the range $14 < \log N_{\rm HI} < 16$. The lower limit is independent of initial cloud density, profile, virialization epoch, depth of the potential well, or geometry of the collapse. Allowing for a cutoff in the radiation field shortward of the He II edge, as may occur if QSOs cannot complete the ionization of helium in the Intergalactic Medium by $z < 5$, permits Doppler parameters as small as $b = 20-25\,{\rm km\,s^{-1}}$. Thus the distribution of Doppler parameters may serve as a probe of the spectral shape of the metagalactic radiation field. A mild correlation of Doppler parameter with neutral hydrogen column density is found in several models for systems with $\log N_{\rm HI} \lesssim 13.5$, with Doppler parameters occurring as low as $b \gtrsim 20\,{\rm km\,s^{-1}}$ for $\log N_{\rm HI} \sim 13$. No lines with $b < 15\,{\rm km\,s^{-1}}$ are found in any of the models computed. No evidence is found for significant distortions from Doppler profiles in any of the spherical models for lines with column densities in the range $13.3 < \log N_{\rm HI} < 16$, suggesting that even supersonic internal velocities may not be detectable in the profiles. Internal velocities may be detected if both the H I and He II Ly$\alpha$ lines may be measured in the clouds. The hydrogen and helium $b$-values will concentrate on a two-branched curve, a thermal branch, for which the helium to hydrogen $b$-value ratio is 0.5, and a second velocity branch with a ratio near unity arising from the internal cloud motions. By contrast, significant profile distortions are found in slabs of sufficient size prior to collapse and may be visible in existing high resolution QSO spectra.

[1]Present address: Department of Astronomy and Astrophysics, University of Chicago, 5640 Ellis Avenue, Chicago, IL 60637.

## 1. Introduction

The last few years have witnessed a considerable refinement in the observational studies of the Ly$\alpha$ forest, detected in the spectra of high redshift QSOs. While the pioneering survey of Sargent et al. (1980) and succeeding efforts over the following decade focused on the density evolution of the absorption lines with redshift and their distribution in equivalent widths, recent high resolution spectroscopy of the forest has permitted accurate measurements to be obtained of the neutral hydrogen column densities and velocity widths of the absorbers through the fitting of Voigt line-profiles to the absorption features (Atwood, Baldwin, & Carswell 1985, Carswell et al. 1987, Pettini et al. 1990, Carswell et al. 1991, Rauch et al. 1992,1993). Although line-blending and systematic errors in the line-fitting of low column density systems continue to plague these determinations, a few features of the Ly$\alpha$ forest distribution have become apparent. It has been established that the distribution in column density is nearly a power law, $dN/dN_{\rm HI} \propto N_{\rm HI}^{-1.7}$, for $\log N_{\rm HI} > 13$ (e.g., Carswell et al. 1991). A single power law, however, does not adequately describe the distribution in detail. On the basis of the published data for 4 QSOs observed at high resolution, Meiksin & Madau (1993a,b) reported a significant deficit in Ly$\alpha$ forest absorbers with $\log N_{\rm HI} > 15$ compared to the extrapolation of the single power-law distribution obtained by fitting to the more abundant lower column density systems. Augmenting the data set with an additional 6 QSOs, Petitjean et al. (1993) examined the distribution in column densities of absorbers over the range $13.7 < \log N_{\rm HI} < 21.8$. They similarly reported structure in the Ly$\alpha$ forest distribution, suggesting a possible break at $\log N_{\rm HI} \sim 14$, below which the distribution may flatten. In addition they report a strong flattening in the distribution at $\log N_{\rm HI} \sim 15.5$. The high resolution measurements have also demonstrated that the Doppler parameters $b$ for the absorbers lie predominantly in the range $20\,{\rm km\,s^{-1}} < b < 40\,{\rm km\,s^{-1}}$. Pettini et al. (1990) report $b$-values lower than $15\,{\rm km\,s^{-1}}$ and a correlation between Doppler parameter and column density for $12.7 < \log N_{\rm HI} < 15$ systems in their $6.5\,{\rm km\,s^{-1}}$ resolution spectrum of Q2206–199, but Rauch et al. (1993) have disputed the reality of both claims. In a re-analysis of the Q2206–199 spectrum, Rauch et al. (1993) argue that the low Doppler parameters and the $b - N_{\rm HI}$ correlation may be artifacts of the line identification and fitting procedure, confirming the conclusion reached earlier by Rauch et al. (1992) in their study of Q0014 + 813.

Despite the recent observational advances, progress in the theoretical understanding of the structure and origin of the Ly$\alpha$ forest absorbers has not been as rapid. There is at present no definitive model for the systems, although a multitude of possibilities has been suggested. Perhaps one of the most promising is the minihalo model of Ikeuchi (1986) and Rees (1986). Rees particularly emphasized the naturalness of the Ly$\alpha$ forest in cosmologies dominated by cold dark matter (CDM). The baryons, photoionized by the metagalactic UV radiation field, are trapped by the collapsing dark matter halos and reach densities sufficiently high that the neutral hydrogen column densities of the clouds exceed $10^{13}\,{\rm cm^{-2}}$. Implications of the minihalo model have been explored by Ikeuchi & Norman (1987), Bond, Szalay, & Silk (1988), Ikeuchi, Murakami, & Rees (1988, 1989), and Murakami & Ikeuchi (1990, 1993).

These papers have demonstrated that the minihalo model may give rise to the observed global distribution of column densities and the measured values for the Doppler parameters in the Ly$\alpha$ forest (e.g., Murakami & Ikeuchi 1993). They do not, however, account for features in the column density distribution or for the observed *range* of the Doppler parameters. The column densities and Doppler parameters of the clouds will depend on their origin and evolution. Thus, a more complete test of the minihalo hypothesis must place the clouds in a cosmological context. With the exception of Bond et al. (1988), previous models for the clouds were constructed in static minihalos. In this paper, the approach of Bond et al. is followed in evolving the clouds from linear primordial perturbations. It is of particular interest to determine the lowest $b$-values expected in the minihalo model and whether a correlation between $b$-value and column density may arise. It is therefore necessary to solve for the cloud temperature self-consistently. While the core of a cloud, where the hydrogen density exceeds $10^{-4}\,{\rm cm^{-3}}$ and the neutral hydrogen column density exceeds $10^{14}\,{\rm cm^{-2}}$, will be very nearly in thermal equilibrium, the density in the outer layers will be too low to have sufficient time to equilibrate. Instead, the outer region of a cloud will retain a partial memory of its initial post-photoionization temperature. The temperature will be further modulated by internal velocities, from both collapse and expansion. Indeed, it is the lower column density systems ($N_{\rm HI} < 10^{14}\,{\rm cm^{-2}}$), that show the lowest $b$-values and a possible correlation of the $b$-values with column density. A central goal of this paper is to compute the effect of deviations from thermal and hydrostatic equilibrium on the column densities and Doppler parameters in the minihalo model.

Both spherical and slab symmetries are treated. The former describes nearly spherical collapse of the dark matter, while the latter describes clouds generated by caustics in the dark matter distribution. While the slab model has been advocated by Charlton, Salpeter, & Hogan (1993), the primary purpose of treating the model here is to examine the effect of geometry on the cloud properties.

The remainder of the paper is organized as follows. In §2, the details of the models are described. The results of various runs for different initial conditions and radiation fields are presented in §3. A summary and discussion are provided in §4. The Appendix contains an algorithm for solving the time-dependent ionization equations.

## 2. Model

The absorbers are evolved from linear perturbations starting at $z = 200$ for spherical systems and $z = 1000$ for planar. The effects of photoionization, collisional ionization and excitation, optically thin radiative recombination losses (case A) from hydrogen and helium, dielectronic recombination losses from helium, collisional ionization and excitation losses, thermal bremsstrahlung, and Compton cooling from scattering off the microwave background are included. A total helium to hydrogen number ratio of $\psi = 1/12$ is assumed throughout. The fluid equations solved are

$$\frac{\partial\rho}{\partial t} + \boldsymbol{\nabla}\cdot(\mathbf{u}\rho) = 0 \; , \tag{1}$$

$$\frac{\partial\mathbf{u}}{\partial t} + \mathbf{u}\cdot\boldsymbol{\nabla}\mathbf{u} + \frac{1}{\rho}\boldsymbol{\nabla}p = -\boldsymbol{\nabla}\phi \; , \tag{2}$$

and

$$\frac{\partial}{\partial t}\log\frac{p}{\rho^\gamma} + \mathbf{u}\cdot\boldsymbol{\nabla}\log\frac{p}{\rho^\gamma} = (\gamma-1)\mathcal{H}/p \; . \tag{3}$$

The photoionization heating rate minus the free-bound and free-free radiative losses, collisional losses, and Compton cooling losses per unit volume are represented by $\mathcal{H}$ (e.g., Osterbrock 1989, Weymann 1965). The photoionization cross-sections are adopted from Osterbrock (1989). The radiative recombination rates for H I and He II are taken from Seaton (1959) and from Aldrovandi & Péquignot (1973) for He I, including dielectronic recombination. The collisional rates are taken from Black (1981). The fits of Shapiro & Kang (1987) to the gaunt factors tabulated by Spitzer (1978) are used for the thermal bremsstrahlung losses from hydrogen and helium. A cosmic background radiation temperature of $2.74(1+z)$ is assumed for Compton cooling. The equation of state for the gas is $p = \rho k_{\rm B}T/\mu m_{\rm H}$, where the mean molecular weight of the gas is $\mu = (1+4\psi)/[1+\chi_{\rm H\,II} + \psi(1+\chi_{\rm He\,II} + 2\chi_{\rm He\,III})]$, $\chi_i$ denoting the ionization fraction of species $i$, $m_{\rm H}$ the mass of a hydrogen atom, and $k_{\rm B}$ Boltzmann's constant. The hydrogen ionization fraction is initially assumed to be at its postrecombination value of $2\times10^{-4}$ (Peebles 1968).

The photoionization field is switched on abruptly at a given redshift. To compute the temperature of the clouds once the field turns on, it is necessary to solve for the time-dependent ionization of the clouds, which initially occurs on a much shorter timescale than the recombination time. Assuming ionization equilibrium immediately after the radiation is switched on was found to underestimate the correct cloud temperature by as much as a factor of two for the fiducial model described below. To solve the ionization equations, the numerical approach of Mathews (1965) for a pure hydrogen H II region is generalized to include helium. The algorithm is described in the appendix.

The average energy deposited into the gas per photoionization is

$$\frac{\int_{\nu_1}^{\infty} h_{\rm P}(\nu-\nu_1)s_\nu U_\nu d\nu/\nu}{\int_{\nu_1}^{\infty} s_\nu U_\nu d\nu/\nu}\ , \qquad (4)$$

(Spitzer 1978), where $\nu_1$ is the frequency at the ionization edge, $s_\nu$ is the photoionization cross-section, $U_\nu$ is the specific energy density of the total ionizing radiation field, and $h_{\rm P}$ is Planck's constant. This energy will be shared by all the constituents of the photoionized gas. For a radiation energy density that may be approximated as decreasing monotonically with increasing frequency, the maximum temperature fully photoionized gas may reach is given by

$$k_{\rm B}T \simeq \frac{1}{3}\frac{1.1+8.4\psi}{2+3\psi}{\rm Ryd}\ , \qquad (5)$$

or $T\simeq 4.2\times10^4$ K. The corresponding Doppler parameter is $b=(2k_{\rm B}T/m_{\rm H})^{1/2}\simeq 26\,{\rm km\,s^{-1}}$. Once radiative recombinations become important, the gas will approach ionization equilibrium. Because the ionization equilibrium timescale of a given species is very short, on the order of the product of the neutral fraction and the total recombination time, ionization equilibrium will be maintained thereafter. The gas may then reach thermal equilibrium. In Figure 1, the Doppler parameter is shown for various densities assuming ionization and thermal equilibrium in a radiation field of intensity $J_\nu = 10^{-21}(\nu/\nu_L)^{-1}\,{\rm ergs\,cm^{-2}\,s^{-1}\,Hz^{-1}\,sr^{-1}}$, where $\nu_L$ is the frequency at the H I Lyman edge. For $n_{\rm H}\gtrsim 4\times10^{-4}\,{\rm cm^{-3}}$, the dominant cooling mechanism is the collisional excitation of neutral hydrogen, resulting in a trend of increasing temperature with decreasing density. The trend is broken at densities near $10^{-4}\,{\rm cm^{-3}}$ by Compton cooling off the cosmic background radiation. By coincidence, the average density of a typical Ly$\alpha$ forest cloud of diameter 20 kpc and neutral hydrogen column density $10^{14}\,{\rm cm^{-2}}$ is also $\sim 10^{-4}\,{\rm cm^{-3}}$. The Doppler parameters range from 15 to $35\,{\rm km\,s^{-1}}$, in good agreement with the measured distribution for the Ly$\alpha$ forest (Rauch et al. 1992).

The agreement, however, must at least in part be fortuitous, because the timescale to reach thermal equilibrium for $n_{\rm H}\lesssim 10^{-4}$ is longer than the age of the universe over the relevant redshifts. The baryons in systems with column densities smaller than $10^{14}\,\mathrm{cm}^{-2}$ must retain a partial memory of their initial postphotoionization temperature. The temperature will be further influenced by the contraction of the gas in deep potential wells and by free-expansion in shallow ones. Moreover, the fluid motions themselves will contribute to the measured line widths. These are not negligible effects. As Bond et al. (1988) emphasized, the minihalo model is an essentially dynamic model. Halos are continually collapsing, and the gas detected as Ly$\alpha$ forest absorbers will in general not be static. Indeed, the contribution of nonthermal motions to the line-widths is indicated by the broad range in Doppler parameters: half of the measured $b$-values (e.g., Rauch et al. 1992), exceed the maximum value of $35\,\mathrm{km\,s}^{-1}$ shown in Figure 1 for gas in ionization and thermal equilibrium at high redshift. While gas which contracts to high densities may give rise to thermal Doppler parameters as low as $\sim 15\,\mathrm{km\,s}^{-1}$, it is unclear whether dynamically stable clouds of such densities may originate from cosmological linear perturbations, or whether they may do so without large concomitant temperatures or velocities in the outer layers which will force the measured Doppler parameters to higher values. These issues will be addressed in this paper.

The radiation field is assumed to be either QSO dominated or galaxy dominated. In the QSO dominated case, the estimate of Meiksin & Madau (1993b) for a constant-comoving density of QSOs is adopted. Specifically, a metagalactic flux of $J_\nu = 2.8\times10^{-22}(\nu/\nu_L)^{-\alpha}\,\mathrm{ergs\,cm^{-2}\,s^{-1}\,Hz^{-1}\,sr^{-1}}$ is used. The temperatures of the clouds are not very sensitive to fluxes near this value, and so the column densities obtained may be scaled approximately inversely with an alternative flux. The spectral index $\alpha$ is scaled according to $\alpha = 0.2 + 1.2/(1+z)$ as an approximation to the hardening of the spectrum by Lyman limit systems and the Ly$\alpha$ forest (Madau 1992). The flux shortward of the He II edge is highly uncertain. According to the estimate of Madau (1992), the flux may be reduced by a factor of $\sim 25$ relative to the H I Lyman edge at high redshifts because of absorption by the Ly$\alpha$ forest and Lyman limit systems. The amount of absorption, however, is not well established. The absorption results largely from systems with H I column densities greater than $10^{15}\,\mathrm{cm}^{-2}$, where the statistics are particularly uncertain. The deficit of Ly$\alpha$ forest clouds reported by Meiksin & Madau (1993b) would result in a much smaller optical depth, based on Madau's (1992) estimate of the contribution from the Ly$\alpha$ forest assuming no deficit. To judge the effect of a break, the flux was reduced at the He II edge compared to the He I edge by the factor $4.5(1+z)$, which is a rough fit to the results of Madau (1992), yielding a central density 8% (37%) higher and central temperature 5% (12%) lower at $z = 5$ (2), for the fiducial spherical model described below compared to the assumption of no break. A break thus has only a minor effect on the thermal and dynamical structure of a cloud. Of greater impact is the absence of flux shortward of the He II edge altogether. This may occur if the He III regions photoionized by QSOs have not yet fully percolated by $z < 5$, or if the He II column density through a cloud grows sufficiently large that the He III in the core recombines. The latter may arise in a cloud for which the H I column density exceeds $10^{16}\,\mathrm{cm}^{-2}$ if the metagalactic flux at the He II edge is as small as suggested by the models of Madau (1992). Radiative transfer effects are not included in this paper, so that the $b$ values may be somewhat smaller than reported here for such systems. Two cases are considered to bracket the uncertainty in the He II flux. In the first no break is assumed, while in the second flux shortward of the He II edge is cut-off completely. For the galactic radiation field, a flat spectrum of intensity $10^{-21}\,\mathrm{ergs\,cm^{-2}\,s^{-1}\,Hz^{-1}\,sr^{-1}}$, cut-off at the He II edge (Madau 1991), and with a turn-on redshift of $z_{\rm ion} = 10$, is added to the QSO spectrum.

Only perturbations virializing or forming caustics after the epoch of reionization are considered. Systems which collapse earlier will have much more complicated histories because of rapid central cooling and the likely formation of stars at their centers. Only systems with halo circular velocities $v_c$ in the range $20\,\mathrm{km\,s^{-1}} < v_c < 60\,\mathrm{km\,s^{-1}}$ are computed. The baryons in more massive halos quickly collapse to form neutral cores and less massive halos are unable to retain their baryons after photoionization. The baryonic mass in these halos is comparable to the estimated total mass for the Ly$\alpha$ forest. Following the approach of Narayan & White (1988) for top-hat spherical perturbations, it may be shown that the fraction of mass captured in halos with circular velocities in the range $v_{c1} < v_c < v_{c2}$ at redshift $z$ is $F(v_{c1}, v_{c2}; z) = \mathrm{erf}[1.68(1+z)/(2^{1/2}\Delta_2)] - \mathrm{erf}[1.68(1+z)/(2^{1/2}\Delta_1)]$, where the *rms* mass fluctuation $\Delta(r_0)$ in a top-hat perturbation of comoving radius $r_0$ is related to the halo circular velocity through $v_c \simeq 1.67(1+z)^{1/2}H_0 r_0$. For a CDM spectrum with bias factor in the range $1.5 - 2.5$, 10%–20% of the mass collapses into halos with circular velocities of $20\,\mathrm{km\,s^{-1}} < v_c < 60\,\mathrm{km\,s^{-1}}$ in the redshift range $2 < z < 5$. Adopting $\Omega_b h_{50}^2 = 0.05 \pm 0.01$ for the nucleosynthesis constraint on the total baryon density from Walker et al. (1991), the baryonic density contribution of the Ly$\alpha$ forest to the Einstein-deSitter critical density would be $\Omega_{\mathrm{Ly}\alpha} \sim 0.005 - 0.01$. Although the spherical assumption is highly approximate, this value is comparable to recent estimates of $\Omega_{\mathrm{Ly}\alpha} \sim 0.002 - 0.05$ for the forest (Meiksin & Madau 1993b, Peebles 1993).

A diffuse Intergalactic Medium (IGM) density of $\Omega_D = 0.04$ or $0.06$ is taken, based on the nucleosynthesis constraint. A case with $\Omega_D = 0.02$ is also tried, allowing for loss to collapsed structures. The main conclusions are not very sensitive to these choices. Unless noted otherwise, the clouds are assumed to be photoionized by QSOs at $z_{\rm ion} = 6$.

The fluid equations are solved in either spherical or slab symmetry. In the case of spherical symmetry, to compute the gravity the dark matter is treated as collapsing shells, following Bond et al. (1988). Equally spaced shells are evolved from an initially linear perturbation. Each shell is brought to rest at its virial radius. The evolution of shell $j$ is parameterized according to

$$r_j = r_{V,j}(1 - \cos\theta_j); \qquad t = \frac{r_{V,j}}{v_{c,j}}(\theta_j - \sin\theta_j)\,, \tag{6}$$

where $r_j$ is the radius of the shell at time $t$, $r_{V,j}$ is the virial radius of the shell, and $v_{c,j}$ is the circular velocity $(GM_j/r_{V,j})^{1/2}$ of the shell, of mass $M_j$, at the time of virialization. Following Bond et al. (1988), the initial linear perturbations are assumed to have a Gaussian density profile. This prescription is found to yield a core with a $\rho \propto r^{-2.25}$ halo, as expected for a cold collapse, to good approximation. The initial profiles are parameterized by the epoch of virialization $z_V$ for the equivalent top-hat perturbation of initial overdensity identical to the initial Gaussian central overdensity and by the corresponding circular velocity $v_c$ at the virialized position of the radius at which the density in the initial profile has declined by a factor of $\mathrm{e}^{-1}$. This value is close to the somewhat radius-dependent halo circular velocity after virialization. For a top-hat perturbation, the virialization epoch for an initial over-density $\delta_i$ at epoch $z_i$ is given by $1 + z_V = (5/3)[8/(9\pi + 6)]^{2/3}\delta_i(1 + z_i) \simeq 0.63\delta_i(1 + z_i)$. A top-hat perturbation was also tried. Similar results were obtained for the clouds, although their edges tended to be more sharply defined. The gradual density decline of the Gaussian perturbation is likely to be more realistic. The dark matter contribution to the gravitational acceleration at $r$ is

$$\frac{\partial\phi}{\partial r} = -\frac{GM(r)}{r^2} , \tag{7}$$

where $M(r)$ is the mass of dark matter within $r$ found by cubic interpolation on a grid of dark matter shells. The self-gravity of the baryons is included for the gas evolution, but a constant baryon to dark matter fraction is assumed for the baryonic contribution to the gravity acting on the shells. This is done so that the shell evolution may be computed directly at each timestep. Because of the small relative density of the baryons, this approximate treatment of the dark matter will not much influence the solutions.

For slab symmetry it is necessary to integrate the evolution of the dark matter directly because of slab crossing, and the effect of the baryon gravity on the dark matter is included. Prior to slab crossing and when the pressure of the baryons may be neglected, the exact solution for the collapsing slabs is given by

$$x = \xi - b_s(t)\eta(\xi); \qquad v = -\dot{b}_s\eta(\xi); \qquad \rho = \rho_0 a^{-3}\left(1 - b_s\frac{d\eta}{d\xi}\right)^{-1} , \tag{8}$$

(Zel'dovich 1970), where $\eta(\xi)$ describes the initial deformation of the slabs in terms of their Lagrangian positions $\xi$, $x$ is the comoving coordinate of a slab, $v = dx/dt$, and $b_s$ is the growing mode solution of the linear density fluctuation equation $\ddot{b}_s + 2(\dot{a}/a)\dot{b}_s + 3(\ddot{a}/a)b_s = 0$, with solution (Groth & Peebles 1975), $b_s = B(\Omega^{-1} - 1)/B(\Omega_0^{-1} - 1)$ for $\Omega_0 < 1$ where $B(y) = 1 + 3/y + 3(1+y)^{1/2}\log[(1+y)^{1/2} - y^{1/2}]/y^{3/2}$, and $b_s = a$ for $\Omega_0 = 1$, where $a = 1/(1+z)$ is the expansion factor. The value of $\Omega$ at redshift $z$ is related to its value $\Omega_0$ today by $\Omega^{-1} - 1 = a(\Omega_0^{-1} - 1)$. The gravitational force per unit mass acting a proper distance $X = ax$ from the center of slab symmetry is, both before and after shell crossing,

$$\frac{\partial\phi}{\partial X} = \frac{3}{2}\Omega H^2 a\left[\frac{1}{2}\lambda_0 f_{DM} q_{DM}(x,t) + \frac{1}{2}\lambda_0 f_b q_b(x,t) - X\right] , \tag{9}$$

(e.g., Bond et al. 1984), where $f_{DM}$ is the fraction of mass in the universe comprised of dark matter, $q_{DM}(x,t)$ is the fraction of the equal mass dark matter slabs originally distributed out to a comoving distance $\lambda_0/2$ that are within a comoving distance $x$ from the center of symmetry at time $t$, and $f_b$ and $q_b(x,t)$ are the corresponding quantities for the baryons. It was found that using 500 slabs faithfully reproduces the similarity solution of Fillmore & Goldreich (1984) for an initial power-law mass perturbation (their $\epsilon = 0.6$ case) out to the fourth turn-around point. In this paper, the initial perturbation is taken to be sinusoidal, $\eta(\xi) = (\lambda_0/2\pi b_{s,c})\sin(2\pi\xi/\lambda_0)$, where $b_{s,c}$ corresponds to the redshift at the time of caustic formation. It is found that as few as 200 slabs is an adequate number for obtaining an accurate gravitational force acting on the baryons for the sinusoidal perturbation. The transverse components of the slabs are assumed to expand along with the Hubble expansion. Although this assumption neglects the influence of the growing overdensity which will slow the transverse expansion of finite slabs, it is adopted for its contrast to the spherical case and for its simplicity. It is a good approximation for slabs sufficiently flat that the epochs for which their transverse sides collapse are much later than those considered.

The fluid equations are solved by finite-differences, using cubic spline collocation to evaluate the spatial derivatives according to the scheme of Bertschinger (Cioffi, McKee, & Bertschinger 1988), who kindly provided a copy of his code. Artificial viscosity is used to treat shocks. The time-integration is explicit and second order accurate. The code is run in its Lagrangian mode. A typical computation uses 200 fluid zones and slightly more zones for the dark matter, whether spherical shells or slabs. Doubling the number of zones or halving the timesteps is found to change the solutions by less than 1%.

## 3. Results

### 3.1. Spherical Models

In the minihalo model, Ly$\alpha$ clouds will be created continuously as new halos collapse after the epoch of reionization $z_{\rm ion}$. Halos with circular velocities smaller than the sound speed of $10^4$ K photoionized gas will not be able to trap the baryons, while halos that are too deep will result in the collapse of the central regions, which is more relevant to the process of galaxy formation. Similarly, systems which collapse prior to reionization are also likely to give rise to star formation and other effects which will not be considered here. Thus, the range of virialization epochs and halo circular velocities considered is $2 < z_V < z_{\rm ion}$ and $20\,{\rm km\,s^{-1}} < v_c < 60\,{\rm km\,s^{-1}}$. The results for a fiducial model are presented, as well as models based on varying the circular velocity, the redshift of virialization, the baryon density, and the metagalactic radiation field. For the fiducial model, $\Omega_0 = 1$ and $H_0 = 50\,{\rm km\,s^{-1}\,Mpc^{-1}}$ are adopted, and the assumed circular velocity and virialization redshift are $v_c = 41\,{\rm km\,s^{-1}}$ and $z_V = 5.4$.

The results of the models are summarized in Table 1. The H I column density refers to the column density along a line-of-sight passing through the cloud center at several redshifts $z$. The range in Doppler parameters, given in kilometers per second, is for absorption systems along any line-of-sight with $13.3 < \log N_{\rm HI} < 16$. This corresponds to the range within which the Ly$\alpha$ optical depth is greater than unity for $T = 3\times10^4$ K, a typical cloud temperature, and the optical depth at the Lyman edge is less than 0.1. The latter limit ensures that radiative transfer effects for the lines-of-sight examined are negligible in those clouds which develop optically thick cores. The range also corresponds to the range in the measured column density values for the Ly$\alpha$ forest reported in the published literature. The circular velocity $v_c$ of the virialized halo is in kilometers per second.

### *3.1.1. Fiducial Model*

The evolution of a perturbation with $z_V = 5.4$ and $v_c = 41\,\mathrm{km\,s^{-1}}$, shown in Figure 2, exhibits several representative features of the minihalo model. The central H I column density ranges between $15.3 < \log N_{\rm HI} < 16.1$ for $2 < z < 5$. Initially, the column density decreases with time because of the large expansion velocity within the cloud at $z = 5$. The initial outflow is a photoionization-driven wind and is a generic feature of clouds which collapse after the photoionization epoch. The wind brings the ratio of the central baryon to dark matter density down to 0.012 at $z = 5$, about a third of the cosmic value of 0.04. The projected radius within which the line-of-sight column density exceeds $10^{14}\,\mathrm{cm^{-2}}$ is 30 kpc, in agreement with the minimum cloud diameter of $10 - 20h_{50}^{-1}$ kpc inferred by Smette et al. (1992) from observations of the lensed QSO UM673 ($H_0 = 50h_{50}\,\mathrm{km\,s^{-1}\,Mpc^{-1}}$). The core of the cloud is nearly in thermal equilibrium. By $z = 3$, hydrostatic equilibrium has nearly been established in the cloud center, although a small adjusting accretion flow remains. The central density ratio at this time has increased somewhat to 0.016. The remnants of the wind have almost vanished. The cloud continues to grow through cosmological accretion. At $z = 2$, the H I column density is still increasing slowly with time and the size of the core is contracting as the cloud continues to approach equilibrium. A sharply defined core develops by this time for $\log N_{\rm HI} > 15\,\mathrm{cm^{-2}}$, a feature common to the other models with such large column densities. Deviations from thermal equilibrium become pronounced for $r > 20$ kpc, as shown in Figure 2. The central accretion results in a diminished density beyond this radius. Because the time to reach thermal equilibrium at these densities exceeds the age of the universe, the temperature stays nearly constant, thus falling short of the equilibrium value for the lower density. The ratio of baryon to dark matter central density is 0.031, and the projected radius corresponding to a column density of $10^{14}\,\mathrm{cm^{-2}}$ is 10 kpc.

The characteristic equilibrium cloud radius may be estimated by noting that the gravitational potential is dominated by the dark matter. An isothermal gas cloud of temperature $T$ and mean mass per particle $\mu m_{\rm H}$ in hydrostatic equilibrium in a uniform density gravitating background has a Gaussian density profile $\rho = \rho_0 \exp(-r^2/2r_c^2)$, with core radius $r_c$. The central dark matter density for a collapsing spherical perturbation after virialization is $\rho_{DM} = (9/8)(3\pi + 2)^2(1 + z_V)^3\rho_c(0)$. The equilibrium core radius is then given by $r_c = [(2k_{\rm B}T/\mu m_{\rm H})/(8\pi G\rho_{DM}/3)]^{1/2} \simeq 27(1 + z_V)^{-3/2}T_4^{1/2}h_{50}^{-1}$ kpc, where $T_4$ is the cloud temperature in units of $10^4$ K. For $z_V = 5.4$ and $T_4 = 3$, the core radius is $r_c \simeq 2.9h_{50}^{-1}$ kpc. For the baryonic contribution to the gravity to remain negligible, the hydrogen number density must be small compared to $n_{\rm H} \approx 3.1 \times 10^{-4}(1 + z_V)^3 h_{50}^2\,\mathrm{cm^{-3}}$. Larger densities will lead to an unstable core because of the gravitational instability of self-gravitating isothermal spheres. This critical number density may be expressed as a critical H I column density. The column density along a line-of-sight passing through the cloud at a projected distance $b_\perp$ from the center is $N_{\rm HI}(b_\perp) = 2\int d(r^2 - b_\perp^2)^{1/2} n_{\rm HI} = \pi^{1/2} n_{\rm HI}(0) r_c \exp(-b_\perp^2/r_c^2)$, where $n_{\rm HI}(r)$ is the H I density distribution in the cloud. Assuming ionization equilibrium in a radiation field of intensity $J_{-21} \times 10^{-21}\,\mathrm{ergs\,cm^{-2}\,s^{-1}\,Hz^{-1}\,sr^{-1}}$ at the Lyman edge gives a critical central H I column density

of

$$N_{\rm HI,crit} \simeq 2.1\times10^{15}(1+z_V)^{9/2}J_{-21}^{-1}T_4^{-0.25}h_{50}^3\,{\rm cm}^{-2}\ . \qquad (10)$$

For $T_4 = 3$ and $J_{-21} = 1$, this corresponds to a column density of $2\times10^{17} < N_{\rm HI}h_{50}^{-3} < 5\times10^{18}\,{\rm cm}^{-2}$, for $2 < z_V < 5$. These column densities correspond to an optical depth near unity at the Lyman edge. Although the analysis starts to lose validity at these optical depths, the increasing density and decreasing radiation field reduce the core temperature and enhance the instability. The cores would also become increasingly neutral as they collapsed. These conditions are favorable to star formation. Since systems with H I column densities exceeding $10^{17}\,{\rm cm}^{-2}$ show metal absorption features, the possibility is suggested that at least some of the metal absorbers may arise in the cores of Ly$\alpha$ clouds (Ikeuchi & Norman 1987, Murakami & Ikeuchi 1990,1993).

The Doppler parameter $b$ is derived by fitting a Doppler profile to the actual line profile obtained by integrating along a given line-of-sight through the cloud. (The cloud is sufficiently optically thin that the effects of radiation damping included in the Voigt profile are negligible, so that a Doppler profile fit is adequate.) The optical depth a projected distance $b_\perp$ from the cloud center is given by

$$\tau_\nu(b_\perp) = \int_{r=b_\perp}^{\infty} d(r^2-b_\perp^2)^{1/2} n_{\rm HI} s\phi_0[\nu-\nu_0(1+\frac{v_l}{c})]\ , \qquad (11)$$

where $s$ is the photoionization cross-section at the line center $\nu = \nu_0$, $v_l = \pm(r^2-b_\perp^2)^{1/2}v(r)/r$ is the internal cloud velocity projected along the line-of-sight, and $\phi_0(\nu) = [\pi^{1/2}\Delta\nu_D]^{-1}\exp[-(\nu-\nu_0)^2/(\Delta\nu_D)^2]$ is the Doppler line profile of width $\Delta\nu_D = \nu_0 b/c$ and Doppler parameter $b = (2k_BT/m_H)^{1/2}$. In this expression, the contributions to the integral from the two signs of $v_l$ simply add. The integrated profile is found to be very accurately fitted by a Doppler profile. Little indication is given of the internal velocities of the cloud. The Doppler parameter and column density are found by fitting the computed optical depths to $\tau_\nu = N_{\rm HI}s\phi_0(\nu)$ over frequencies for which $\tau_\nu > {\rm Max}(\tau_{GP}, 0.1)$, where $\tau_{GP}$ is the Ly$\alpha$ optical depth of the IGM at the same redshift. The column densities found in this way are essentially identical to those obtained by direct integration along a line-of-sight through the cloud. The Doppler parameters predominantly occupy the range $25-40\,{\rm km\,s^{-1}}$, as shown in Figure 3, with a slight trend toward lower values for $\log N_{\rm HI} < 13.5$ systems by $z = 2$.

Although internal velocities in the cloud are difficult to discern in the shape of the absorption profile, these motions may be revealed by measuring the corresponding He II Ly$\alpha$ features. The H I and He II Doppler parameters for the model are displayed in Figure 4a. The parameters describe a "Y"-shaped locus. One branch has a slope of 0.5 and corresponds to temperature-dominated $b$-values. A steeper second branch, approaching a slope of unity, shows values dominated by the internal cloud motions. In Figure 4b, the ratio of He II to H I $b$-values is shown as a function of H I column density. The ratio increases toward low column densities at the higher redshifts, indicating the higher internal velocities in the younger systems. Simultaneous measurements of the H I and He II Ly$\alpha$ features would offer a unique opportunity to map the internal dynamics of Ly$\alpha$ clouds.

### *3.1.2. Model Variations*

Decreasing the circular velocity to $v_c = 20\,\mathrm{km\,s^{-1}}$ results in clouds which never fully retain their baryons. The potential well is not sufficiently deep to reverse the initial wind after reionization, as shown in Figure 5. As a consequence, the central H I column density continues to decline with time, from $\log N_{\mathrm{HI}} = 14.5$ at $z = 5$ to $\log N_{\mathrm{HI}} = 13.1$ by $z = 2$. The Doppler parameter for lines with $\log N_{\mathrm{HI}} > 12.7$ is always in excess of $22\,\mathrm{km\,s^{-1}}$, and in excess of $25\,\mathrm{km\,s^{-1}}$ for $\log N_{\mathrm{HI}} > 13.5$. Although the low optical depths of clouds with such small circular velocities will cause them to be missed in Ly$\alpha$ forest surveys, the Ly$\alpha$ "thicket" that they would comprise could nonetheless contain a large fraction of the baryons in the IGM.

The absorption line shapes through the clouds, shown in Figure 6, are found to be very well fitted by Doppler profiles even in the presence of bulk velocities comparable to the internal sound speed of the clouds. The robustness of the Doppler profile may be understood by the following consideration. For the ideal case of a spherical isothermal cloud with a Gaussian density profile $n_{\mathrm{HI}}(r) = n_{\mathrm{HI}}(0)\exp(-r^2/r_c^2)$ and linear velocity field $v(r) = v_*(r/r_c)$, it is straightforward to show that lines-of-sight passing through the cloud exactly preserve a Voigt line profile, with the Doppler parameter $b$ replaced by $b = (2k_{\mathrm{B}}T/m_H + v_*^2)^{1/2}$ in the expression for the Doppler width $\Delta\nu_D$. The preservation of a Doppler profile follows as a special case. Clouds in a shallow potential well will be in near free-expansion. The free-expansion of an isothermal cloud approaches self-similarity with a Gaussian density profile and linear velocity profile (Imshennik 1960). The similarity solutions apply in cylindrical and planar symmetries as well. Thus, it is the deviations from a linear velocity profile that will give rise to distortions from a Doppler profile in the expanding clouds.

Results similar to those for the $v_c = 41\,\mathrm{km\,s^{-1}}$ case are found for a circular velocity of $v_c = 35\,\mathrm{km\,s^{-1}}$. In this case, near hydrostatic equilibrium is reached by $z = 4$ with a central density of $n_{\mathrm{H}} \simeq 3.7 \times 10^{-4}\,\mathrm{cm^{-3}}$, a central column density of $\log N_{\mathrm{HI}} \simeq 14.7$, and an isothermal core with a nearly constant temperature of $T \simeq 3.8 \times 10^4\,\mathrm{K}$. The diameters of $N_{\mathrm{HI}} = 10^{14}\,\mathrm{cm^{-2}}$ systems lie in the range $10 - 30\,\mathrm{kpc}$. The cloud corresponds closely to the hydrostatic minihalo model envisioned by Ikeuchi (1986)and Rees (1986). A minimum Doppler parameter for $\log N_{\mathrm{HI}} > 14$ of $b > 25\,\mathrm{km\,s^{-1}}$ is found.

Systems with circular velocities of $v_c = 50\,\mathrm{km\,s^{-1}}$ have gravitationally unstable cores. At $z = 5$, the central column density is $\log N_{\mathrm{HI}} = 15.8$ and has reached $\log N_{\mathrm{HI}} = 17.2$ by $z = 3$. The optically thin approximation is no longer valid. The required radiative transfer through the core of the cloud is beyond the scope of this paper. As discussed above, the most likely outcome may be a neutral star forming core. The $b$-values all lie above $25\,\mathrm{km\,s^{-1}}$ for lines-of-sight with $13.3 < \log N_{\mathrm{HI}} < 16$.

Varying the IGM density between $\Omega_D = 0.02$ and $0.06$ resulted in $b$-values similar to those of the above models, as did decreasing the virialization redshift to $z_V = 3.5$. At most only small deviations from a Doppler profile arise near line center in the models. Increasing the circular velocity tends to exaggerate the distortion. A model with $z_V = 2.5$ and $v_c = 60\,\mathrm{km\,s^{-1}}$ was computed in an attempt to create a large deviation. At $z = 3$, the best-fit intensity at line center is $0.2I_0$, where $I_0$ is the unabsorbed intensity, while the true intensity is $0.04I_0$. Off center the profiles agree well. This is the most extreme deviation found in the spherical cases computed.

The same model illustrates another feature of the minihalo scenario: a halo need not collapse to give rise to a Ly$\alpha$ forest absorber. By $z = 5$, the potential well is sufficiently deep to create a respectable Ly$\alpha$ feature, with a central column density of $10^{14.3}\,\mathrm{cm^{-2}}$ and Doppler parameter of $34\,\mathrm{km\,s^{-1}}$.

If the IGM was photoionized at $z > 6$, Ly$\alpha$ cloud halos could collapse at higher redshifts, resulting in larger central densities and column densities, lower equilibrium temperatures, and possibly lower Doppler parameters. A model with $v_c = 41\,\mathrm{km\,s^{-1}}$ is computed in an alternative radiation field dominated by galaxies for $z > 6$ and allowed to virialize at $z_V = 8.5$. The galaxy field is turned on at $z_{\rm ion} = 10$. The QSOs are turned on at $z_{\rm ion} = 6$ as in the fiducial model, ionizing the He II . By $z = 5$, the core has begun to collapse, with a central column density of $\log N_{\rm HI} = 16.6$. The projected radius for which the column density exceeds $10^{16}\,\mathrm{cm^{-2}}$ is 1.4 kpc. By $z = 2$, the radius is 2 kpc, while the central column density has climbed to $\log N_{\rm HI} = 17.2$. The projected radius at which $\log N_{\rm HI} = 14$ contracts from 10 kpc at $z = 5$ to 6 kpc at $z = 2$. The minimum Doppler parameter found for $14 < \log N_{\rm HI} < 16$ clouds is $b \sim 24\,\mathrm{km\,s^{-1}}$, not much smaller than the systems collapsing at lower redshift. Such an early collapse is consistent with the known properties of Ly$\alpha$ forest systems.

All the cloud models above have $b$-values exceeding $24\,\mathrm{km\,s^{-1}}$ for lines-of-sight with column densities in the range $14 < \log N_{\rm HI} < 16$. A large number of lower $b$-value systems in this column density range, however, are found in high resolution studies (e.g., Rauch et al. 1992). Even ignoring the controversial systems with $b < 15\,\mathrm{km\,s^{-1}}$, a significant fraction of the Doppler parameters measured lie in the range $20\,\mathrm{km\,s^{-1}} < b < 25\,\mathrm{km\,s^{-1}}$, indicating lower cloud temperatures and internal velocities. While the statistics of the distribution near $b \gtrsim 20\,\mathrm{km\,s^{-1}}$ are still uncertain, it is worthwhile examining how low a temperature may be achieved by cutting off the external radiation field at the He II edge. To date, there are no constraints on the epoch of full ionization of helium. It is possible that helium is singly ionized in the IGM for $z > 2$. Using the approach of Meiksin & Madau (1993b), the epoch of complete helium ionization, where the He III porosity $Q_{\rm HeIII}$ reaches unity, for their model CC (constant comoving QSO density switched on at $z_{\rm ion} = 6$), and $\Omega_D = 0.02$, is $z_{\rm ion}^{\rm HeII} \simeq 5.7$ for their medium QSO spectrum. Allowing for a large clumped component with $\Omega_D = 0.01$ and $\Omega_{{\rm Ly}\alpha} = 0.03$ decreases the He III reionization redshift to $z_{\rm ion}^{\rm HeII} \simeq 3.2$. For a softer spectrum, the ionization epoch may be as recent as $z_{\rm ion}^{\rm HeII} < 2$.

A model with $z_V = 5.4$ and $v_c = 35\,\mathrm{km\,s^{-1}}$ is computed for a QSO radiation field which turns on at $z_{\rm ion} = 6$, but with the spectrum cut-off at the He II edge. At $z = 5$, the central column density is $\log N_{\rm HI} = 15.5$, and $b = 21\,\mathrm{km\,s^{-1}}$ for the corresponding absorption line. The projected radius for which $\log N_{\rm HI} = 14$ is 17 kpc, and $b = 24\,\mathrm{km\,s^{-1}}$. At $z = 3$, the central column density is $\log N_{\rm HI} = 15.8$, with $b = 22\,\mathrm{km\,s^{-1}}$, and it continues to climb with time. The projected radius for which $\log N_{\rm HI} = 14$ is now 11 kpc, and $b = 20\,{\rm km\,s^{-1}}$. Thus, significantly smaller $b$-values may be obtained for $\log N_{\rm HI} > 14$ clouds in the absence of He II ionization.

Since the *Hubble Space Telescope* has detected the Ly$\alpha$ forest at low redshift (Bahcall et al. 1991, Morris et al. 1991), it is of interest to consider the structure of a recently collapsed cloud. A model is computed with a virialization redshift of $z_V = 0.5$ and a circular velocity of $v_c = 41\,{\rm km\,s^{-1}}$. The metagalactic radiation field is scaled according to $(1+z)^{3.5}$ for $z < 1.9$ to match the low redshift estimate of Madau (1992). At $z = 0.05$ the central column density is $\log N_{\rm HI} \simeq 14.1\,{\rm cm^{-2}}$, with a $b$-value of $28\,{\rm km\,s^{-1}}$. The corresponding equivalent width that would be measured is 0.29A, typical of the systems reported by Bahcall et al. (1991). The central hydrogen density is $1.2\times10^{-5}\,{\rm cm^{-3}}$, and the central temperature is $5.4\times10^{4}\,{\rm K}$. The average density over a 30 kpc radius is $10^{-5}\,{\rm cm^{-3}}$. This would produce H$\alpha$ in emission at an intensity of $I_{{\rm H}\alpha} \sim 1\times10^{-24}\,{\rm ergs\,cm^{-2}\,sec^{-1}\,arcsec^{-2}}$, well below current thresholds for detection. The column density exceeds $10^{13.3}\,{\rm cm^{-2}}$ out to $b_\perp \simeq 75\,{\rm kpc}$, so that the cloud is quite extended. The Doppler parameter at this point is $b \simeq 21\,{\rm km\,s^{-1}}$.

### 3.2. Slab Models

The spherical model is only an approximation to the gravitational collapse of minihalos. Caustic formation during the collapse of the dark matter will also give rise to Ly$\alpha$ clouds. Only dark matter candidates with a velocity dispersion negligible compared to the sound speed of the gas (as in CDM-dominated cosmologies), are considered, so that the internal velocity dispersion of the dark matter particles may be neglected. In this case, Ly$\alpha$ clouds may form at the sites of the dark matter caustics. The resulting slabs permit an examination of the effect of geometry on the range of column densities and $b$-values expected in dark matter dominated systems. The results for the slab models are presented in Table 2. The $b$-values are for lines-of-sight with column densities in the range $13.3 < \log N_{\rm HI} < 16$, while the range in column density is for lines-of-sight intercepting the slabs at angles $\theta$ from the normal given by $0.1 < \cos\theta < 1$.

Perturbations with comoving wavelengths ranging from 0.5 to 8 Mpc are considered. A typical case, with a redshift for caustic formation of $z_c = 5.5$, is shown in Figure 7. After the caustic forms and dark matter slabs begin to cross, the gas escapes the weakened central potential well and eventually meets an outer cosmological accretion shock. It is noted that the gas bounded by the accretion shocks is nearly isobaric, with the pressure decreasing from the slab center to the shock front by 20%–50%, depending on epoch. A similar result was previously found for more massive pancakes by Bond et al. (1984). Despite the high velocities attained, no evidence for the bulk motion is found in the absorption line profiles other than for line-broadening. The profiles remain Doppler in shape. As in the spherical case, $b$-values smaller than $25\,{\rm km\,s^{-1}}$ are not found in systems with $\log N_{\rm HI} > 14$.

Varying the wavelength of the perturbation and the redshift for caustic formation yielded very similar results. The smallest $b$-value found in any of the models for $\log N_{\rm HI} > 14$ is $26\,{\rm km\,s^{-1}}$. Cutting off the ionizing flux shortward of the He II edge gave $b$-values as low as $21\,{\rm km\,s^{-1}}$ for such systems, while a value as low as $b \simeq 17\,{\rm km\,s^{-1}}$ was found for systems with $\log N_{\rm HI} < 13.4$ at $z = 2$. Doppler parameters smaller than $15\,{\rm km\,s^{-1}}$ were not found.

A case with $\lambda_0 = 8$ Mpc and $z_c = 2.5$ was computed in order to search for distortions from a Doppler profile. The mass of the system would be in excess of $10^{13}\,{\rm M_\odot}$. The case is similar to that of Miralda-Escudé & Rees (1993). As discussed by McGill (1990), the slabs will produce an absorption feature even prior to collapse. The evolution of the line-width is shown in Figure 8. Not until $z = 2$ does a Doppler profile provide a good fit to the lines. The $z = 3$ profile corresponds nearly to the epoch of Figure 1c in Miralda-Escudé & Rees (1993) and confirms the marked distortion from a Doppler profile they find. The column density giving rise to the feature is $10^{13.9}\,{\rm cm^{-2}}$, close to the best-fit value of $10^{14.0}\,{\rm cm^{-2}}$. The formal best-fit Doppler parameter is $59\,{\rm km\,s^{-1}}$. The peak accretion velocity at this time is $58\,{\rm km\,s^{-1}}$, at a distance of 80 kpc from the cloud center. The corresponding temperature is $1.5\times10^4$ K. The peak temperature in the slab at $z = 2$, after caustic-formation, is $2.3\times10^5$ K. By this time, the infall energy has been converted to thermal energy by an accretion shock and the line profile is Doppler in shape. The column density for a line-of-sight normal to the slab is $4.4\times10^{13}\,{\rm cm^{-2}}$, with a best-fit Doppler parameter of $38\,{\rm km\,s^{-1}}$. The central hydrogen density at this time is $1.2\times10^{-4}\,{\rm cm^{-3}}$. Whether such systems are common depends on the power-spectrum of the primordial density fluctuations. The double-lobed nature of the profile at $z = 3$ could be misidentified as two Ly$\alpha$ systems in a QSO spectrum. It may be interesting to examine the high resolution spectra of QSOs for such flattened features.

Because of the transverse velocity in the clouds, a correlation between column density and Doppler parameter is found for the high column density systems. The degree of the correlation, however, will depend on the details of the transverse motion. The motions are likely to be affected by neighboring systems, as $N$-body simulations show a tendency for flattened dark matter structures to drain into one another. These details cannot be addressed in this paper.

## 4. Discussion

Several features of the models are summarized and discussed in this section.

The structure of the clouds may be divided into three zones: (1) a nearly hydrostatic core in thermal equilibrium, (2) an intermediate dynamic zone out of thermal equilibrium, and (3) an outer accretion layer joining onto the Hubble expansion. Systems with column densities above $10^{14} - 10^{15}\,{\rm cm^{-2}}$ will generally arise from lines-of-sight that pass through the cloud cores. Although the baryons will be partly depleted by a postphotoionization wind, the density remains sufficiently high in the core, and may even grow if the wind reverses into a central accretion flow, that thermal equilibrium may be established. The bulk of the Ly$\alpha$ forest, however, is composed of lower column density systems. These will originate from the dynamical intermediate layer. Because of the low density of the layer, the characteristic cooling time of the gas is longer than a Hubble time so that the gas temperature will retain a partial memory of its initial postphotoionization value and may be further influenced by motions within the cloud, both collapse and expansion.

The structure of the clouds has several interesting observational implications. Because of the surrounding accretion zone, the Doppler parameters of the absorption features will lie somewhat above the value corresponding to the equilibrium core temperature. While halos with low circular velocities will have lower accretion velocities, they will retain their baryons less well in the post-photoionization wind, resulting in a lower density hence a higher temperature core, because the temperature of photoionized gas in ionization and thermal equilibrium increases with decreasing density. Decreasing the circular velocity too far will result in a permanent wind. Conversely, systems with high circular velocities will retain their baryons better and have lower temperature cores. Their accretion velocities and accretion temperatures, however, will be higher. If the circular velocity is sufficiently great, the resulting accretion will result in a contraction of the core and possibly in a dynamical collapse if the neutral hydrogen column density reaches $10^{17} - 10^{18}\,\mathrm{cm}^{-2}$, an effective Jeans column density. Because the core will begin to be shielded from the external radiation field at these column densities, it will recombine, accelerating the collapse. Consequently, there is a minimum Doppler parameter for stable spherical systems having $\log N_{\rm HI} > 14$. The limit is found to be $b \sim 25\,\mathrm{km\,s^{-1}}$. It is insensitive to the initial cloud density, profile, virialization epoch (for $z_V > 2$), or depth of the potential well. Moreover, the same limit is found in slab geometry. If the ionization of He II has not been completed in the IGM, the minimum may be as low as $b \sim 20\,\mathrm{km\,s^{-1}}$. These values agree well with the minimum values measured in the Ly$\alpha$ forest at these column densities (Rauch et al. 1992).

Systems that reach central column densities in excess of $10^{15}\,\mathrm{cm}^{-2}$ tend to develop a more sharply defined core, suppressing somewhat the trend of decreasing size with column density. Because the detection probability of an absorber is proportional to its cross-section, the steepening trend detected in the distribution of column densities in the range $13 < \log N_{\rm HI} < 15$ would be similarly suppressed, an effect which may have been detected in the observed flattening of the distribution at $\log N_{\rm HI} \sim 15.5$ (Petitjean et al. 1993).

Absorption features from lines-of-sight passing through the intermediate layer, where thermal equilibrium will not in general be achieved, may have lower $b$-values, reaching values as low as $\sim 20\,\mathrm{km\,s^{-1}}$. As a consequence, a mild trend of decreasing $b$-value with decreasing column density may be expected for $\log N_{\rm HI} \lesssim 13.5$. Although systematic errors resulting from the line-identification and line-fitting procedures may currently preclude a definite detection of such a trend (Rauch et al. 1992,1993), it would serve as a test of the minihalo model that may be made available by larger samples and improved signal-to-noise spectra.

While the computations presented here generally agree well with previous computations in reproducing systems with column densities corresponding to the Ly$\alpha$ forest, they differ in several important regards. Because Bond et al. (1988) assumed a fixed temperature for the clouds, it is not possible to compare against their results for the Doppler parameters. Murakami & Ikeuchi (1993) solved for the temperature explicitly. The structures of their clouds do not reveal the distinctive separation between core and halo found here until the core in their models starts to recombine. This is largely a consequence of their neglect of cosmological infall. They assume instead that the baryons are initially in hydrostatic equilibrium within a static dark matter halo. They also find a lower range for the Doppler parameters, with values occurring as low as $14\,\mathrm{km\,s^{-1}}$ for clouds with column densities in the range $14 < \log N_{\rm HI} < 16$ and $z > 2$. The highest $b$-value they show in this column density range is $25\,\mathrm{km\,s^{-1}}$, corresponding to the lowest value found here. The difference again is partly accounted for by their neglect of the contribution of cosmological infall to the Doppler parameter. It may also, however, in part be a consequence of their assumption of an initial temperature given by thermal equilibrium rather than adopting a postphotoionization temperature, which may result in an underestimate of the contribution of the outer regions of the cloud to the Doppler parameter.

The Doppler profile is found to be extremely robust for the absorption features. Even in the presence of supersonic motion, the distortions are small in the spherical model. The distortions are similarly small for a slab geometry. Searching for profile distortions is thus not expected to prove a successful means of measuring the gas motions in individual clouds. The exception is for the collapse of massive slabs, where the profile flattening found by Miralda-Escudé & Rees (1993) is reproduced. If the He II Ly$\alpha$ line corresponding to the H I lines may be detected, however, the dynamics of the clouds may be revealed. The corresponding Doppler parameters will cluster on a two-branched curve. A thermal branch will arise from $b$-values dominated by the cloud temperature, while a second velocity branch will result from any large motions within the clouds.

A low redshift analog of the high redshift absorption systems is computed, corresponding to the Ly$\alpha$ absorbers recently discovered by the *Hubble Space Telescope*. The cloud structure for systems collapsing late is similar to that of the high redshift clouds.

Both the spherical and slab geometries demonstrate that Ly$\alpha$ features may also arise in systems prior to the gravitational collapse of their dark matter. Thus, the Ly$\alpha$ forest may ultimately be found to trace structures on a variety of scales. If the universe is CDM-dominated, however, it seems likely that a primary component will derive from the baryons in collapsed and collapsing dark matter minihalos.

This work was completed while the author held an International Fellowship from the Natural Sciences and Engineering Research Council of Canada.

## A. Time-Dependent Ionization Algorithm

The rate equations for photoionization, radiatiave recombination, and collisional ionization for a gas of hydrogen and helium are

$$\frac{dx}{dt} = (1-x)G_{\rm HI} - n_{\rm H}[x + \psi(1+z-y)][x\alpha_{\rm HI} - (1-x)\gamma_{\rm HI}] \ , \tag{A1}$$

$$\frac{dy}{dt} = -yG_{\rm HeI} + n_{\rm H}[x + \psi(1+z-y)][(1-y-z)\alpha_{\rm HeI} - y\gamma_{\rm HeI}] \ , \tag{A2}$$

and

$$\frac{dz}{dt} = (1-y-z)G_{\rm HeII} - n_{\rm H}[x + \psi(1+z-y)][z\alpha_{\rm HeII} - (1-y-z)\gamma_{\rm HeII}] \ , \tag{A3}$$

where $x$ denotes the fraction of ionized hydrogen, $y$ the fraction of neutral helium, $z$ the fraction of doubly ionized helium, $\psi$ the ratio of helium to hydrogen number densities, and $n_{\rm H}$ the total number density of hydrogen. The photoionization, radiative recombination, and electron collisional ionization rates for species $i$ ($i$ = H I , He I , or He II ) are denoted respectively by $G_i$, $\alpha_i$, and $\gamma_i$.

A finite-difference representation of equations (A1)–(A3) is found which is an exact solution in the limits of ionization equilibrium and pure photoionization in a time-independent radiation field, the two limits of primary interest here. This is accomplished by factoring the right-hand-sides of the above equations to obtain

$$\frac{dx}{dt} = -a_x(2)(x - x^+)(x - x^-) \ , \tag{A4}$$

$$\frac{dy}{dt} = a_y(2)(y - y^+)(y - y^-) \ , \tag{A5}$$

and

$$\frac{dz}{dt} = -a_z(2)(z - z^+)(z - z^-) + (y^- - y)G_{\rm HeII} \ . \tag{A6}$$

The following quantities have been defined,

$$x^\pm = \frac{-a_x(1) \pm [a_x(1)^2 - 4a_x(0)a_x(2)]^{1/2}}{2a_x(2)} \ , \tag{A7}$$

$$y^\pm = \frac{-a_y(1) \pm [a_y(1)^2 - 4a_y(0)a_y(2)]^{1/2}}{2a_y(2)} \ , \tag{A8}$$

and

$$z^{\pm} = \frac{-a_z(1) \pm [a_z(1)^2 - 4a_z(0)a_z(2)]^{1/2}}{2a_z(2)} , \tag{A9}$$

where $a_x$, $a_y$, and $a_z$ are given by

$$a_x = \begin{pmatrix} -[G_{\rm HI} + \psi(1+z-y)n_{\rm H}\gamma_{\rm HI}] \\ G_{\rm HI} - n_{\rm H}\gamma_{\rm HI} + \psi(1+z-y)a_x(2) \\ n_{\rm H}(\alpha_{\rm HI} + \gamma_{\rm HI}) \end{pmatrix} , \tag{A10}$$

$$a_y = \begin{pmatrix} n_{\rm H}[x + \psi(1+z)](1-z)\alpha_{\rm HeI} \\ -G_{\rm HeI} - n_{\rm H}[x + \psi(1+z)](\alpha_{\rm HeI} + \gamma_{\rm HeI}) - n_{\rm H}\psi(1-z)\alpha_{\rm HeI} \\ n_{\rm H}\psi(\alpha_{\rm HeI} + \gamma_{\rm HeI}) \end{pmatrix} , \tag{A11}$$

and

$$a_z = \begin{pmatrix} -(1-y^-)G_{\rm HeII} - n_{\rm H}[x + \psi(1-y)](1-y)\gamma_{\rm HeII} \\ G_{\rm HeII} + n_{\rm H}[x + \psi(1-y)](\alpha_{\rm HeII} + \gamma_{\rm HeII}) - n_{\rm H}\psi(1-y)\gamma_{\rm HeII} \\ n_{\rm H}\psi(\alpha_{\rm HeII} + \gamma_{\rm HeII}) \end{pmatrix} . \tag{A12}$$

In ionization equilibrium, $x = x^+$, $y = y^-$, and $z = z^+$. Defining $\chi_x = x^+ - x$, $\chi_y = y^- - y$, and $\chi_z = z^+ - z$, equations (A4)–(A6) are solved to second order accuracy in $\Delta t$ by

$$\chi_x^{n+1} = \chi_x^n \exp(-\Delta\tau_x) + \frac{x^{+,n+1} - x^{+,n}}{\Delta\tau_x}[1 - \exp(-\Delta\tau_x)] , \tag{A13}$$

$$\chi_y^{n+1} = \chi_y^n \exp(\Delta\tau_y) + \frac{y^{-,n+1} - y^{-,n}}{\Delta\tau_y}[\exp(\Delta\tau_y) - 1] , \tag{A14}$$

and

$$\begin{aligned} \chi_z^{n+1} = \quad & \chi_z^n \exp(-\Delta\tau_z) + \tfrac{z^{+,n+1} - z^{+,n}}{\Delta\tau_z}[1 - \exp(-\Delta\tau_z)] + \\ & \tfrac{1}{2}\tfrac{y^{-,n+1} - y^{-,n}}{\Delta\tau_y}[1 - \exp(-\Delta\tau_y)]\exp(-\Delta\tau_z)G_{\rm HeII}^n\Delta t - \\ & \begin{cases} \frac{1}{2}\chi_y^{n+1}(G_{\rm HeII}^{n+1} + G_{\rm HeII}^n)\Delta t, & \Delta\tau_y + \Delta\tau_z = 0; \\ G_{\rm HeII}^{n+1}\chi_y^{n+1}\frac{1-\exp(-\Delta\tau_y - \Delta\tau_z)}{\Delta\tau_y + \Delta\tau_z}\Delta t + \frac{1}{2}\chi_y^{n+1}(G_{\rm HeII}^{n+1} - G_{\rm HeII}^n)\Delta t, & \Delta\tau_y + \Delta\tau_z \neq 0, \end{cases} \end{aligned} \tag{A15}$$

where a superscript $n$ denotes evaluation at time $t^n$, $\Delta t = t^{n+1} - t^n$, $\Delta\tau_x = (1/2)[a_x^{n+1}(2)(x^{n+1} - x^{-,n+1}) + a_x^n(2)(x^n - x^{-,n})]\Delta t$, $\Delta\tau_y = (1/2)[a_y^{n+1}(2)(y^{n+1} - y^{+,n+1}) + a_y^n(2)(y^n - y^{+,n})]\Delta t$, and $\Delta\tau_z = (1/2)[a_z^{n+1}(2)(z^{n+1} - z^{-,n+1}) + a_z^n(2)(z^n - z^{-,n})]\Delta t$. In the absence of radiative recombination and collisional ionization, equations (A13)– (A15) are exact solutions of the photoionization equations for a time-independent radiation field. Since the photoionization time is short compared to the time for the radiation field to change for Ly$\alpha$ clouds, the integration is still nearly exact when recombinations and collisions are negligible. The timestep was chosen sufficiently small to ensure that the solutions were accurate to a few percent or better in test cases where radiative recombinations were significant while the ionization fractions were still out of equilibrium. The equations are implicit in $\chi_x^{n+1}$, $\chi_y^{n+1}$, and $\chi_z^{n+1}$ and may be solved by iteration.

Table 1. Spherical Models

| $\Omega_D$ | $z_{\rm ion}^{\rm HI}$ | $z_{\rm ion}^{\rm HeII}$ | $z_V$ | $v_c$ | $z$ | $\log N_{\rm HI}$[a] | $b$[b] |
|---|---|---|---|---|---|---|---|
| 0.04 | 6 | 6 | 5.4 | 20 | 5 | 14.5 | 26–36 |
| | | | | | 4 | 13.9 | 26–44 |
| | | | | | 3 | 13.5 | 25–26 |
| | | | | | 2 | 13.1 | ... |
| 0.04 | 6 | 6 | 5.4 | 35 | 5 | 15.1 | 26–42 |
| | | | | | 4 | 14.7 | 27-40 |
| | | | | | 3 | 14.5 | 25–28 |
| | | | | | 2 | 14.4 | 24–30 |
| 0.04 | 6 | 6 | 5.4 | 41 | 5 | 15.4 | 26–44 |
| | | | | | 4 | 15.3 | 27–43 |
| | | | | | 3 | 15.5 | 26–31 |
| | | | | | 2 | 16.1 | 23–32 |
| 0.04 | 6 | 6 | 5.4 | 50 | 5 | 15.8 | 25–45 |
| | | | | | 4 | 16.3 | 25–47 |
| | | | | | 3 | (17.2) | 25–34 |
| | | | | | 2 | (18.2) | 27–37 |
| 0.02 | 6 | 6 | 5.4 | 41 | 5 | 14.8 | 26–31 |
| | | | | | 4 | 14.6 | 26–28 |
| | | | | | 3 | 14.7 | 22–30 |
| | | | | | 2 | 15.1 | 26–32 |
| 0.06 | 6 | 6 | 5.4 | 35 | 5 | 15.5 | 26–42 |
| | | | | | 4 | 15.1 | 27–42 |
| | | | | | 3 | 14.8 | 27–37 |
| | | | | | 2 | 14.7 | 25–30 |
| 0.04 | 6 | 6 | 3.4 | 52 | 5 | 14.5 | 29–51 |
| | | | | | 4 | 14.4 | 29–56 |
| | | | | | 3 | 14.2 | 31–49 |
| | | | | | 2 | 14.7 | 23–35 |
| 0.04 | 6 | 6 | 2.5 | 60 | 5 | 14.3 | 34–58 |
| | | | | | 4 | 14.2 | 34–65 |
| | | | | | 3 | 13.9 | 34–58 |
| | | | | | 2 | 14.1 | 23–33 |
| 0.04 | 10 | 6 | 8.5 | 41 | 5 | 16.6 | 24–33 |
| | | | | | 4 | 16.7 | 23–31 |
| | | | | | 3 | 16.9 | 24–32 |
| | | | | | 2 | 17.2 | 24–34 |

Table 1—Continued

| $\Omega_D$ | $z_{\rm ion}^{\rm HI}$ | $z_{\rm ion}^{\rm HeII}$ | $z_V$ | $v_c$ | $z$ | $\log N_{\rm HI}$[a] | $b$[b] |
|---|---|---|---|---|---|---|---|
| .04 | 6 | 0 | 5.4 | 35 | 5 | 15.5 | 21–36 |
| | | | | | 4 | 15.4 | 21–31 |
| | | | | | 3 | 15.8 | 19–25 |
| | | | | | 2 | 16.4 | 19–27 |
| 0.04 | 6 | 6 | 0.5 | 41 | 0.5 | 13.2 | ··· |
| | | | | | 0.3 | 13.5 | 20–23 |
| | | | | | 0.05 | 14.1 | 21–29 |
| | | | | | 0 | 14.2 | 22–29 |

[a]H I column density along line-of-sight through cloud center. Values in parentheses refer to values optically thick at the Lyman edge, and so are not accurately computed.

[b]Doppler parameter $b$ ( $\rm km\,s^{-1}$) range for lines-of-sight with $13.3 < \log N_{\rm HI} < 16$. Free expansion in the external layers may result in large $b$-values and a slightly increasing column density with impact parameter through the cloud. In these cases, the largest $b$-value, when it exceeds all previous $b$-values, before the increase in column density is given in order to provide a more representative expression of the range.

Table 2. Slab Models

| $\Omega_D$ | $z_{\rm ion}^{\rm HI}$ | $z_{\rm ion}^{\rm HeII}$ | $z_c$ | $\lambda_0$ | $z$ | $\log N_{\rm HI}$[a] | $b$[b] |
|---|---|---|---|---|---|---|---|
| 0.04 | 6 | 6 | 5.5 | 0.5 | 5 | 13.5–14.4 | 22–47 |
| | | | | | 4 | 13.1–14.0 | 24–58 |
| | | | | | 3 | 12.7–13.6 | 45–65 |
| | | | | | 2 | 12.7–13.0 | ... |
| 0.04 | 6 | 6 | 5.5 | 1 | 5 | 13.9–14.8 | 25–44 |
| | | | | | 4 | 13.5–14.4 | 22–53 |
| | | | | | 3 | 13.1–14.0 | 22–58 |
| | | | | | 2 | 12.6–13.5 | 42–62 |
| 0.04 | 6 | 6 | 5.5 | 2 | 5 | 14.3–15.2 | 28–40 |
| | | | | | 4 | 14.0–14.9 | 27–43 |
| | | | | | 3 | 13.6–14.5 | 24–46 |
| | | | | | 2 | 13.0–14.0 | 22–47 |
| 0.04 | 6 | 6 | 5.5 | 4 | 5 | 14.6–15.6 | 32–44 |
| | | | | | 4 | 14.4–15.4 | 31–35 |
| | | | | | 3 | 14.0–15.0 | 28–35 |
| | | | | | 2 | 13.4–14.4 | 25–36 |
| 0.04 | 6 | 6 | 2.5 | 1 | 5 | 13.6–14.4 | 25–104 |
| | | | | | 4 | 13.8–14.2 | 22–101 |
| | | | | | 3 | 12.9–14.1 | 20–35 |
| | | | | | 2 | 12.6–13.3 | 75–79 |
| 0.04 | 6 | 6 | 2.5 | 2 | 5 | 13.8–14.6 | 27–164 |
| | | | | | 4 | 13.6–14.4 | 22–124 |
| | | | | | 3 | 13.2–14.1 | 20–90 |
| | | | | | 2 | 12.8–13.7 | 27–59 |
| 0.04 | 6 | 6 | 2.5 | 8 | 5 | 14.1–15.2 | 56–578 |
| | | | | | 4 | 13.9–14.9 | 35–321 |
| | | | | | 3 | 13.7–14.6 | 34–117 |
| | | | | | 2 | 13.6–14.5 | 38–44 |

Table 2—Continued

| $\Omega_D$ | $z_{\rm ion}^{\rm HI}$ | $z_{\rm ion}^{\rm HeII}$ | $z_c$ | $\lambda_0$ | $z$ | $\log N_{\rm HI}$[a] | $b$[b] |
|---|---|---|---|---|---|---|---|
| .04 | 10 | 6 | 7.5 | 2 | 5 | 14.0–14.9 | 27–37 |
| | | | | | 4 | 13.5–14.5 | 25–42 |
| | | | | | 3 | 13.1–14.1 | 23–45 |
| | | | | | 2 | 12.6–13.5 | 33–45 |
| 0.04 | 6 | 0 | 5.5 | 2 | 5 | 14.5–15.3 | 24–35 |
| | | | | | 4 | 14.1–15.1 | 23–35 |
| | | | | | 3 | 13.8–14.7 | 20–35 |
| | | | | | 2 | 13.2–14.1 | 17–35 |

[a]H I column density range for line-of-sight impact parameter angles $\theta$ from the vertical of $0.1 < \cos\theta < 1$.

[b]Doppler parameter $b$ ($\rm km\,s^{-1}$) range for lines-of-sight with $13.3 < \log N_{\rm HI} < 16$.

---

This preprint was prepared with the AAS LaTeX macros v3.0.

Fig. 1.— Thermal Doppler parameter $b_{\rm th}$ as a function of hydrogen density $n_{\rm H}$ for a gas of cosmic abundance in ionization and thermal equilibrium, assuming an ionizing flux $J_\nu = 10^{-21}(\nu/\nu_L)^{-1}\,{\rm ergs\,cm^{-2}\,s^{-1}\,Hz^{-1}\,sr^{-1}}$, where $\nu_L$ is the frequency at the Lyman edge. Shown at redshifts $z = 5$ (*solid line*), $z = 4$ (*long-dashed line*), $z = 3$ (*short-dashed line*), and $z = 2$ (*dotted line*). The turn-over for low densities is a result of Compton cooling off the cosmic background radiation. The thermal equilibrium time exceeds the age of the universe for $n_{\rm H} \lesssim 10^{-4}\,{\rm cm^{-3}}$. Physical systems will deviate from the curves shown at lower densities by an amount which may depend on their history. Half of the measured Doppler parameters exceed the maximum equilibrium value of $35\,{\rm km\,s^{-1}}$ for $z > 2$.

Fig. 2.— Evolution of the hydrogen density $n_{\rm H}$, the temperature $T$, the difference $\Delta T_{\rm eq} = T - T_{\rm eq}$ of the temperature from the thermal equilibrium value $T_{\rm eq}$ for the same density as the cloud density (*thin lines*), and the velocity $v$, as a function of cloud radius $r$ for the fiducial model, with virialization redshift $z_V = 5.4$ and circular velocity $v_c = 41\,{\rm km\,s^{-1}}$. Also shown is the neutral hydrogen column density $N_{\rm HI}$ as a function of projected distance $b_\perp$ from the cloud center. An initial photoionization-driven wind ultimately reverses to an inner accretion flow as the cloud approaches hydrostatic equilibrium. Shown at redshifts $z = 5$ (*solid line*), $z = 4$ (*long-dashed line*), $z = 3$ (*short-dashed line*), and $z = 2$ (*dotted line*).

Fig. 3.— Doppler parameter $b$ as a function of $N_{\rm HI}$ for the fiducial model. Shown at redshifts $z = 5$ (*solid line*), $z = 4$ (*long-dashed line*), $z = 3$ (*short-dashed line*), and $z = 2$ (*dotted line*). A mild correlation at low column densities arises at late epochs from a dynamical layer outside the cloud core.

Fig. 4.— (a) He II Doppler parameter $b_{\rm HeII}$ as a function of the H I Doppler parameter $b_{\rm HI}$ for the fiducial model, revealing the dynamic nature of the cloud. The two branches correspond to a thermal branch of slope 0.5 and a velocity branch of slope 1. Shown at redshifts $z = 5$ (*solid line*), $z = 4$ (*long-dashed line*), $z = 3$ (*short-dashed line*), and $z = 2$ (*dotted line*). (b) Ratio of $b_{\rm HeII}$ to $b_{\rm HI}$ as a function of H I column density for the model. Higher ratios, hence greater internal velocities, are apparent in the younger systems.

Fig. 5.— Same as Fig. 2, but for the case $z_V = 5.4$ and $v_c = 20\,{\rm km\,s^{-1}}$. A postphotoionization wind persists because of the small depth of the halo potential.

Fig. 6.— Absorption line profile (*solid line*), and best-fit Doppler profile (*dashed line*), through the cloud center for the case of Fig. 5. Shown at $z = 5$, 4, 3, and 2, decreasing upward. The Doppler profile provides an excellent fit to the lines despite the large gas velocities.

Fig. 7.— Evolution of hydrogen density $n_{\rm H}$, temperature $T$, and velocity $v$, as a function of the proper distance $X$ from the plane of collapse, for the slab model with comoving wavelength $\lambda_0 = 2\,{\rm Mpc}$ and epoch of caustic formation $z_c = 5.5$. Also shown is the neutral hydrogen column density $N_{\rm HI}$ as a function of $\cos\theta$, where $\theta$ is the angle of the line-of-sight from the vertical. Shown at redshifts $z = 5$ (*solid line*), $z = 4$ (*long-dashed line*), $z = 3$ (*short-dashed line*), and $z = 2$ (*dotted line*).

Fig. 8.— Absorption line profile (*solid line*), and best-fit Doppler profile (*dashed line*), for a vertical line-of-sight through the slab for the slab model with comoving wavelength $\lambda_0 = 8\,{\rm Mpc}$ and redshift of caustic formation $z_c = 2.5$. Shown at $z = 4$, 3, and 2, decreasing upward. A marked flattening in the profile is visible shortly before a caustic develops. After the caustic forms, the infall energy thermalizes and the line profile becomes Doppler in shape.